\documentclass[epj]{svjour}
\usepackage{graphics}
\usepackage{amsmath,amssymb}
\usepackage{graphicx}
\usepackage[dvips]{epsfig}
\usepackage{subfigure}

\newcommand{\pol}[1]{\mathaccent"017E{#1}}

\begin{document}
\title{Measurement of the analyzing powers in $\boldsymbol{\pol{p}d}$ elastic and $\boldsymbol{\pol{p}n}$
       quasi-elastic scattering at small angles}
\titlerunning{Measurement of the analyzing powers in ${\pol{p}d}$ elastic and ${\pol{p}n}$
       quasi-elastic scattering at small angles}

\author{S.~Barsov\inst{1}\thanks{Email: barsov\!\_sg@pnpi.nrcki.ru}\and
Z.~Bagdasarian\inst{2,3}\and
S.~Dymov\inst{4,5}\and
R.~Gebel\inst{3}\and
M.~Hartmann\inst{3}\and
A.~Kacharava\inst{3}\and
I.~Keshelashvili\inst{3}\and
A.~Khoukaz\inst{6}\and
V.~Komarov\inst{4}\and
P.~Kulessa\inst{7}\and
A.~Kulikov\inst{4}\and
A.~Lehrach\inst{3}\and
N.~Lomidze\inst{2}\and
B.~Lorentz\inst{3}\and
G.~Macharashvili\inst{2}\and
D.~Mchedlishvili\inst{2,8}\and
S.~Merzliakov\inst{4,3}\and
S.~Mikirtychyants\inst{1}\and
M.~Nioradze\inst{2}\and
D.~Prasuhn\inst{3}\and
F.~Rathmann\inst{3}\and
D.~Schr\"{o}er\inst{6}\and
V.~Serdyuk\inst{3}\and
V.~Shmakova\inst{4}\and
R.~Stassen\inst{3}\and
H.~Str\"oher\inst{3}\and
M.~Tabidze\inst{2}\and
A.~T\"{a}schner\inst{6}\and
S.~Trusov\inst{9,10}\and
D.~Tsirkov\inst{4}\and
Yu.~Uzikov\inst{4,10,11}\and
Yu.~Valdau\inst{1,3}\and
C.~Wilkin\inst{12}\thanks{Email: c.wilkin@ucl.ac.uk (corresponding author)} }

\institute{
St.\ Petersburg Nuclear Physics Institute, NRC Kurchatov Institute, RU-188350
Gatchina, Russia
\and High Energy Physics Institute, Tbilisi State University, GE-0186
Tbilisi, Georgia
\and Institut f\"ur Kernphysik, Forschungszentrum J\"ulich, D-52425
 J\"ulich, Germany
\and Laboratory of Nuclear Problems, JINR, RU-141980 Dubna, Russia
\and University of Ferrara and INFN, I-44100 Ferrara, Italy
\and Institut f\"ur Kernphysik, Universit\"at M\"unster, D-48149 M\"unster,
Germany
\and H.~Niewodnicza\'{n}ski Institute of Nuclear Physics PAN, PL-31342
Krak\'{o}w, Poland
\and SMART$|$EDM-Lab, Tbilisi State University, GE-0179 Tbilisi, Georgia
\and Institut f\"ur Kern- und Hadronenphysik, Forschungszentrum Rossendorf,
D-01314 Dresden, Germany
\and Department of Physics, M.~V.~Lomonosov Moscow State University,
RU-119991 Moscow, Russia
\and Dubna State University, RU-141980 Dubna, Russia
\and Physics and Astronomy Department, UCL, London WC1E 6BT, UK}

\date{Received: \today / Revised version:}
\abstract{The analyzing powers in proton-deuteron elastic and proton-neutron
quasi-elastic scattering have been measured at small angles using a polarized
proton beam at the COSY storage ring incident on an unpolarized deuterium
target. The data were taken at 796~MeV and five higher energies from 1600~MeV
to 2400~MeV. The analyzing power in $pd$ elastic scattering was studied by
detecting the low energy recoil deuteron in telescopes placed symmetrically
in the COSY plane to the left and right of the beam whereas for $pn$
quasi-elastic scattering a low energy proton was registered in one of the
telescopes in coincidence with a fast scattered proton measured in the ANKE
magnetic spectrometer. Though the experiment explores new domains, the
results are consistent with the limited published information.
} 
\maketitle

\section{Introduction}
\label{Introduction}

The nucleon-nucleon ($NN$) interaction is of great importance in any study of
hadronic processes at intermediate energies. At such energies a full set of
amplitudes may be extracted using a phase-shift analysis but this is
obviously dependent on the availability of a reliable experimental data base.
Proton-proton elastic scattering has been extensively studied in many
laboratories worldwide, including at the COoler SYnchrotron (COSY) of the
Forschungs\-zentrum
J\"ulich~\cite{ALB1997,ALB2004,ALT2000,ALT2005,BAU2003,BAU2005,BAG2014,MCH2016}.
The wealth of spin-dependent quantities measured has allowed the extraction
of $NN$ phase shifts in the isospin I=1 channel up to almost
3000~MeV~\cite{ARN2000,ARN2007}. The situation is far less promising for the
isoscalar channel where the much poorer neutron-proton data base only permits
the I = 0 phase shifts to be evaluated up to at most 1300~MeV, but with
significant ambiguities above about 800~MeV.

Small angle neutron-proton elastic scattering has been studied at COSY over
recent years by measuring the interaction of a deuteron beam with a hydrogen
target~\cite{MCH2013,ADL2014}. However, in this case the maximum beam energy
at COSY is about 1150~MeV/nucleon. To go higher in energy, where $np$ data
are very scarce, measurements have to be performed using a proton beam
incident on a deuterium target.

The differential cross section~\cite{MCH2016} and analyzing
power~\cite{BAG2014} in proton-proton elastic scattering have been studied at
COSY using the ANKE magnetic spectrometer. Despite the ANKE acceptance and
experimental capabilities for investigating $p n$ elastic scattering becoming
much less favourable as the beam energy increases, it was considered a
priority for the ANKE collaboration to contribute to the $pn$ elastic data
base above 1500~MeV by measuring the proton analyzing power.

The elements of ANKE that were used in this experiment are described in
sect.~\ref{Experiment}. These are the forward detector, in which fast protons
were measured, and the silicon tracking telescopes (STT) that were used to
measure low energy protons and deuterons. Since the results were obtained
with a polarized proton beam, its preparation and measurement were integral
to the success of the proposal. However, the experiment was carried out just
after the measurement of the analyzing power in proton-proton elastic
scattering~\cite{BAG2014} using the same beam so that the presentation in
sect.~\ref{Beam} can be relatively brief.

Proton-deuteron elastic scattering could be cleanly identified and measured
by detecting the deuteron in one of the STT without the use of the forward
detector. As described in sect.~\ref{pd}, with two STT placed symmetrically
(left and right) around the target to form a two-arm polarimeter, the proton
analyzing power in $pd$ elastic scattering could be measured in a way that is
completely analogous to the analyzing power measurement in $pp$ elastic
scattering~\cite{BAG2014}. Though the measurements at 796~MeV are consistent
with published results to within experimental uncertainties~\cite{IRO1983},
there are no other data at 1600~MeV and above with which to make comparisons.

The measurement of the analyzing power in proton-neutron quasi-elastic
scattering, which is the subject of sect.~\ref{NN}, is much more challenging.
Though the $pd\to ppn$ reaction can be selected by measuring one fast proton
in the forward detector and a slow proton in an STT, there is then the
difficulty of identifying quasi-free elastic $pn$ collisions and avoiding
regions where the $NN$ final state interaction (FSI) is very strong. Ideally,
the contamination from these effects would be studied with the help of a full
reaction model but, in its absence, one has to resort to a more empirical
approach.

The $pn$ FSI, which can lead in particular to the reformation of a deuteron,
decreases fast with the momentum transfer, as does the $pd$ elastic
differential cross section itself. Furthermore, the contribution from
quasi-free scattering on the neutron in the deuteron is enhanced in regions
where the ``spectator'' proton momentum is small compared to the overall
momentum transfer ($q$). Both these features can be exploited by making
appropriate kinematic cuts. This empirical approach was tested successfully
on data taken at 796~MeV.

Unlike proton-deuteron elastic scattering, the left-right symmetry is lost
when measuring analyzing powers with a combination of an STT and the forward
detector. One is then left with a one-arm polarimeter that relies on
measurements of the intensities of the polarized beams as well as their
polarizations. Nevertheless, the results obtained are consistent with the
limited available published information. Our conclusions are drawn in
sect.~\ref{Conclusions}.

\section{Experimental setup}
\label{Experiment}

\begin{figure}[h]
\includegraphics[width=1.0\linewidth]{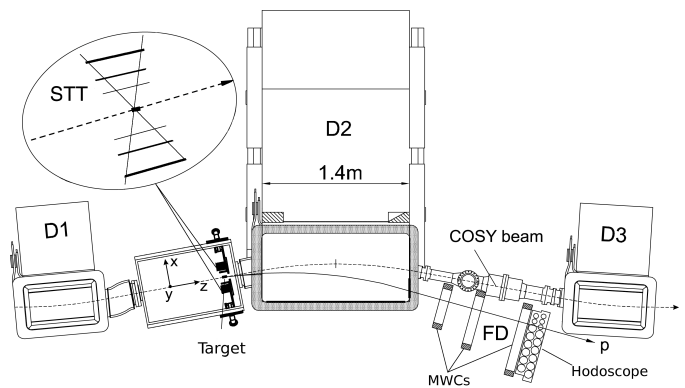}
\caption{The ANKE spectrometer setup (top view), showing the positions of
the deuterium cluster-jet target, the silicon tracking
telescopes (STT), and the forward detector (FD).
}
\label{ankesetup}
\end{figure}

The experiment was carried out using the ANKE magnetic
spectrometer~\cite{BAR2001} positioned inside the COSY storage
ring~\cite{MAI1997} of the  Forschungszentrum J\"ulich. Although the facility
sketched in Fig.~\ref{ankesetup}  was equipped with other elements, the only
detectors used in this experiment were the forward detector (FD) and the
silicon tracking telescopes (STT)~\cite{SCH2003}.

Fast protons arising from small-angle proton-deuteron elastic scattering or
quasi-free elastic scattering on the constituent nucleons were measured in
the FD in the range $4^{\circ}-10^{\circ}$ in laboratory polar angle
($\theta_{\rm lab}$) and $160^{\circ}-200^{\circ}$ in azimuthal angle
($\phi$). The forward detector comprises a set of multiwire proportional and
drift chambers and a two-plane scintillation hodoscope. In addition to their
use for triggering, the scintillators were also needed to measure the energy
losses required for particle identification~\cite{DYM2004}.

The two STT were installed in the COSY plane symmetrically inside the vacuum
chamber to the left and right of the beam at distances of 3~cm from the
deuterium cluster-jet target, which had a diameter of about
10~mm~\cite{KHO1999}. Each telescope consists of three position-sensitive
silicon layers of 70~$\mu$m, 300~$\mu$m, and 5~mm thickness and, in this
configuration, covered laboratory polar angles $75^{\circ} < \theta_{\rm lab}
< 140^{\circ}$. The acceptances of the STT in azimuth of $\pm 30^{\circ}$
were centred at $\phi=0^{\circ}$ and $\phi=180^{\circ}$ on the left and the
right sides, respectively.

Protons and deuterons were clearly identified by the $dE-E$ technique when
they passed through the first layer and were stopped in the second or third
layer of an STT. These conditions are realized for protons with kinetic
energies between 2.5 and 30~MeV and for deuterons between 3.5 and 40~MeV. The
momenta of these low energy protons and deuterons were determined using the
position information from the first and the second layers and their total
energy loss. The relative positions of the silicon detectors in the first and
the second layers were directly measured in the laboratory with a precision
of $\pm0.1$~mm. The front-end electronics of the STT provided the
self-triggering signal from the second layers (\emph{STT-trigger}).

\section{Polarized proton beam}
\label{Beam}

The ANKE experiment used a vertically polarized beam incident on an
unpolarised target and the preparation of the beam and the measurement of its
polarization were carried out in common with the studies of the analyzing
powers in proton-proton elastic scattering~\cite{BAG2014}. H$^-$ ions, with
either spin up $(\uparrow)$ or down $(\downarrow)$, were supplied by the
polarized ion source. These were then accelerated to 45~MeV in the cyclotron
JULIC before being stripped of their electrons and injected into the COSY
ring~\cite{EVE1993}. The sign of the polarization was flipped at every beam
injection at the beginning of the acceleration cycle. The polarization of the
injected beam was optimized using a low energy polarimeter in the injection
beam line to COSY~\cite{CHI2006}. In both spin modes, source polarizations of
about 0.93 were achieved and the difference between their values was measured
to be smaller than the statistical uncertainty of 1\%.

In a strong-focusing synchrotron, such as COSY, resonances can lead to losses
of polarization of a proton beam during acceleration. In order to compensate
for these effects, adiabatic spin-flip was used to overcome the imperfection
resonances and tune-jumping to deal with the intrinsic ones~\cite{LEH2003}.
The beam polarization after acceleration was measured using the EDDA detector
as a polarimeter. This detector, originally equipped with a polarized
hydrogen target, had been used to measure the analyzing power in elastic
proton-proton scattering at larger angles over almost the whole COSY energy
range~\cite{ALT2000,ALT2005}. By studying further the scattering of polarized
protons on C and CH$_2$ targets, it was possible to deduce the quasi-free
analyzing power of carbon, where the necessary calibration standard was
provided by the EDDA $p\pol{p}$ data~\cite{WEI2000}.

The simplified version of the EDDA detector that was used in the present
experiment was equipped with a 7~$\mu$m diameter carbon fibre target that
could be moved in and out of the beam. The polarimeter, which had been
calibrated during the EDDA data-taking periods against the full detector
setup, consists of 29 pairs of half-rings placed to the left and right of the
beam. The left-right asymmetry of counts is determined for each pair of
half-rings, thus providing a dependence on the polar angle $\theta_{\rm lab}$
while averaging over the azimuthal angle $\phi$ in every half-ring. The
systematic uncertainty of the measurements was estimated to be
$3\%$~\cite{WEI2000}.

The experiment was carried out at six proton kinetic energies, $T_p=796$,
1600, 1800, 1965, 2157, and 2368~MeV. Cycles of 180~s or 300~s duration were
used, with the last 20~s of each cycle being reserved for the measurement of
the beam polarization with the EDDA polarimeter~\cite{ANK2014}. Mean values
of the beam polarizations determined from the EDDA data at the six energies
are given in Table~\ref{polar}. It should be noted that the values correspond
to half the difference between spin-up and spin-down data because the
simplified variant of the EDDA detector does not allow the determination of
the polarization for each spin mode individually. The changes in sign reflect
the number of spin flips required to pass through the imperfection
resonances. Since each of the six beams was prepared independently by the
COSY crew, the magnitude of the polarization need not decrease monotonically
as further resonances are crossed.

\begin{table}[h!]
\caption{The mean values of beam polarizations $P$ determined with the EDDA
polarimeter averaged over all the data at the beam energy $T_p$ in MeV. The
changes in the sign of $P$ are due to the spin flips induced when passing
through the imperfection resonances. Though the statistical errors shown are
small, there are $3\%$ systematic uncertainties~\cite{WEI2000}.
\label{polar} \vspace{3mm} }%
\centering \scriptsize{
\begin{tabular}{|c|c|c|c|c|c|c|}
\hline
$T_p$&796&1600&1800&1965&2157&2368\\
\hline
$P$       & \,0.511\,&\,0.378\,&\,$-0.476$\,&\,$-0.508$\,&\,$-0.513$\,&\,$0.501$\,\\
          & $\pm0.001$ & $\pm0.001$ & $\pm0.003$ & $\pm0.005$ & $\pm0.005$ & $\pm 0.004$\\
\hline
\end{tabular}
}
\end{table}

\section{Analyzing power in proton-deuteron elastic scattering}
\label{pd}

\begin{figure}[h]
\includegraphics[width=0.9\linewidth]{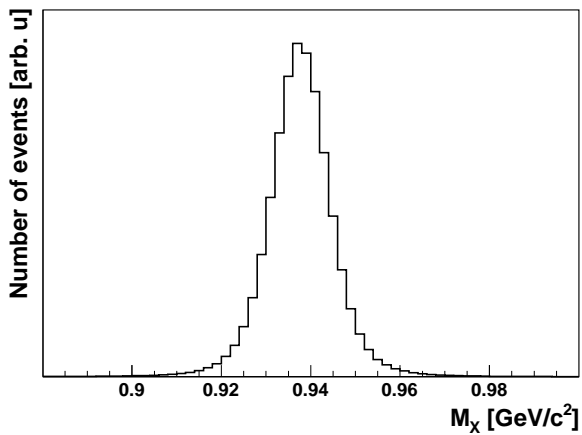}
\caption{The missing-mass distribution for the $pd\to dX$ reaction at
$T_p=796$~MeV where the slow deuteron was detected in the left-side STT.
Since no information from the forward detector was used here, the widths of the
distributions are almost independent of beam energy. For the same reason, the
results obtained from the two STT yield, of course, indistinguishable
missing-mass distributions.}\label{Mx_dSTT1}
\end{figure}

Elastic proton-deuteron scattering was the only source of low energy
deuterons that fell within the angular acceptance of the STT. This reaction
can therefore be reliably identified by just evaluating the information
provided by STT. For this purpose, events were recorded using the
\emph{STT-trigger}, which requested a minimal energy deposit in the second
layer of either of the two STT telescopes. The deuterons were then easily
selected from energy loss measurements in the silicon layers. As a
consequence, it is not surprising that the missing-mass distributions in the
$pd \rightarrow dX$ reaction measured in either STT showed only clear peaks,
well centred at the proton mass, with very little background, as illustrated
in Fig.~\ref{Mx_dSTT1} using data from one STT at a beam energy of 796~MeV.
The positions of these peaks were independent of the deuteron kinetic energy
($T_d$).  In both STT the peaks had the same widths of 15.6~MeV/c$^2$
(\emph{FWHM}), as averaged over the total $T_d$ range. The widths increased
significantly with decreasing $T_d$, due to small angle scattering of the
deuterons in the first and second layers of the STT.

From the numbers of deuterons detected in the left ($L_d$) and right ($R_d$)
telescopes during each acceleration cycle, the asymmetry of $\pol{p} d$
elastic scattering was evaluated for each pair of successive cycles with beam
polarizations up and down, using the cross-ratio method~\cite{OHL1973}, which
eliminates first-order systematic errors. It was carefully studied in
Ref.~\cite{ANK2014} for the $\pol{p} p$ elastic data, which were taken under
similar conditions but with the hydrogen cluster target. None of the cycles
were used twice and for each beam energy the asymmetry over the data-taking
period was quite stable and the result constant to within statistical
uncertainties. These data thus allowed us to detect if there were any
variation of the beam polarization cycle by cycle. The $L_d/R_d$ ratio, which
was calculated for each cycle, was constant within statistical errors for
each of the two spin modes. This indicates that, not only the beam
polarization, but also the acceptances of the STT, were quite stable during
measurements at all beam energies. Less than 1\% of cycles at each beam
energy were found to show any significant deviation of the $L_d/R_d$ ratio
from its average value. The data from these cycles were not considered in the
subsequent analysis.

\begin{table*}[ht]
\centering
\begin{tabular}{|c|c|c|c|c|c|c|}
\hline
$\Theta_{cm}$ & $A_y^{p}(pd:796)$ & $A_y^{p}(pd:1600)$& $A_y^{p}(pd:1800)$& $A_y^{p}(pd:1965)$& $A_y^{p}(pd:2157)$& $A_y^{p}(pd:2368)$\\
degrees  & & & & & & \\
\hline
\phantom{1}4.5&       ---     &       ---       &        ---  &$0.100\pm0.011$&$0.086\pm0.007$&$0.082\pm0.005$\\
\phantom{1}5.5&       ---     &$0.115\pm0.003$&$0.123\pm0.002$&$0.097\pm0.002$&$0.092\pm0.003$&$0.091\pm0.002$\\
\phantom{1}6.5&       ---     &$0.106\pm0.002$&$0.132\pm0.001$&$0.108\pm0.002$&$0.105\pm0.002$&$0.099\pm0.002$\\
\phantom{1}7.5&$0.257\pm0.012$&$0.127\pm0.001$&$0.143\pm0.001$&$0.123\pm0.002$&$0.113\pm0.002$&$0.113\pm0.002$\\
\phantom{1}8.5&$0.268\pm0.004$&$0.139\pm0.001$&$0.154\pm0.001$&$0.132\pm0.001$&$0.126\pm0.002$&$0.118\pm0.002$\\
\phantom{1}9.5&$0.293\pm0.003$&$0.146\pm0.001$&$0.165\pm0.001$&$0.141\pm0.002$&$0.133\pm0.002$&$0.125\pm0.002$\\
10.5          &$0.316\pm0.003$&$0.156\pm0.002$&$0.181\pm0.001$&$0.152\pm0.002$&$0.142\pm0.003$&$0.132\pm0.002$\\
11.5          &$0.340\pm0.003$&$0.167\pm0.002$&$0.186\pm0.001$&$0.162\pm0.002$&$0.149\pm0.003$&$0.136\pm0.003$\\
12.5          &$0.358\pm0.003$&$0.169\pm0.002$&$0.191\pm0.001$&$0.165\pm0.002$&$0.155\pm0.003$&$0.143\pm0.003$\\
13.5          &$0.378\pm0.002$&$0.179\pm0.002$&$0.200\pm0.002$&$0.167\pm0.003$&$0.161\pm0.004$&$0.147\pm0.004$\\
14.5          &$0.392\pm0.003$&$0.182\pm0.002$&$0.207\pm0.002$&$0.174\pm0.003$&$0.160\pm0.005$&$0.136\pm0.005$\\
15.5          &$0.410\pm0.003$&$0.174\pm0.003$&$0.202\pm0.002$&$0.160\pm0.004$&$0.153\pm0.006$&$0.133\pm0.008$\\
16.5          &$0.415\pm0.003$&$0.184\pm0.003$&$0.197\pm0.003$&$0.162\pm0.005$&$0.150\pm0.011$&$0.150\pm0.011$\\
17.5          &$0.423\pm0.003$&$0.172\pm0.004$&$0.197\pm0.004$&$0.156\pm0.013$&       ---     &       ---     \\
18.5          &$0.425\pm0.004$&$0.188\pm0.005$&       ---     &       ---     &       ---     &       ---     \\
19.5          &$0.427\pm0.004$&$0.193\pm0.010$&       ---     &       ---     &       ---     &       ---     \\
20.5          &$0.434\pm0.004$&       ---     &       ---     &       ---     &       ---     &       ---     \\
21.5          &$0.434\pm0.005$&       ---     &       ---     &       ---     &       ---     &       ---     \\
22.5          &$0.430\pm0.005$&       ---     &       ---     &       ---     &       ---     &       ---     \\
23.5          &$0.425\pm0.007$&       ---     &       ---     &       ---     &       ---     &       ---     \\
24.5          &$0.411\pm0.008$&       ---     &       ---     &       ---     &       ---     &       ---     \\
25.5          &$0.405\pm0.009$&       ---     &       ---     &       ---     &       ---     &       ---     \\
26.5          &$0.424\pm0.011$&       ---     &       ---     &       ---     &       ---     &       ---     \\
27.5          &$0.398\pm0.016$&       ---     &       ---     &       ---     &       ---     &       ---     \\
28.5          &$0.406\pm0.090$&       ---     &       ---     &       ---     &       ---     &       ---     \\
\hline
\end{tabular}
\caption{Analysing power $A_y^{p}(pd)$ in $\pol{p}d$ elastic scattering at
six proton kinetic energies marked in the separate columns in MeV.}
\label{tab:Aypd}
\end{table*}

\begin{figure}[h]
\includegraphics[width=1.0\linewidth]{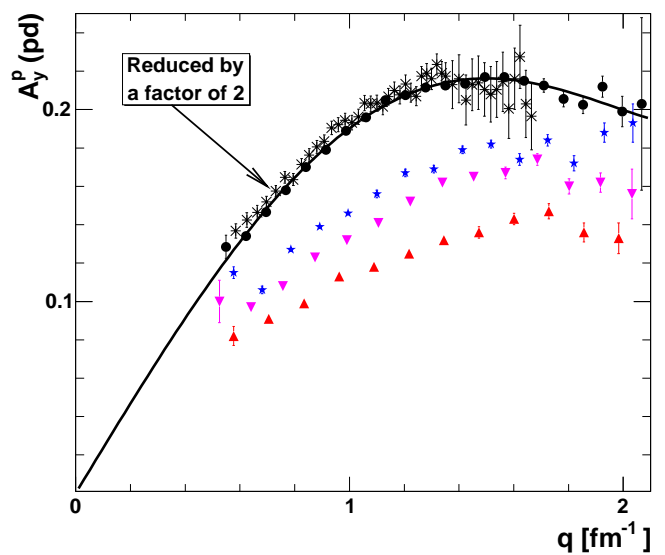}
\caption{The proton analyzing power $A_y^{p}(pd)$ in proton-deuteron elastic
scattering as a function of the momentum transfer $q$ in the centre-of-mass
frame. ANKE data at 796~MeV are shown by closed (black) circles, at 1600~MeV
by (blue) stars, at 1965~MeV by (magenta) inverted triangles, and at 2368~MeV
by (red) triangles. Only statistical errors are shown and in general these are
smaller than the symbol size. The fit of Eq.~(\ref{fit}) to the ANKE data at
796~MeV is shown by the continuous curve. The LAMPF data at this energy are
shown by (black) crosses~\cite{IRO1983}. In order to increase the visibility
of the higher energy points, the results and curve at 796~MeV are reduced by a
factor of two. The numerical values of the ANKE data are to be found in
Table~\ref{tab:Aypd}.}
\label{pdAy_STT}
\end{figure}

The angular dependence of the proton analyzing power in $\pol{p} d$ elastic
scattering was determined from the \emph{STT-trigger} data for all six beam
energies and the results are shown in Fig.~\ref{pdAy_STT} in terms of the
c.m.\ momentum transfer $q$. The numerical values at all six energies are
presented in Table~\ref{tab:Aypd} as a function of the scattering angle
$\Theta_{cm}$. The values of  $\Theta_{cm}$ and $q$ were determined from the
deuteron kinetic energy, which was measured in the STT to much higher
accuracy than the polar angle. The deuteron energy was measured with about a
2\% uncertainty, which would correspond to an uncertainty in $\Theta_{cm}$ of
less than $0.2^{\circ}$ in the angular range below $20^{\circ}$.

On general grounds the proton analyzing power\footnote{We use a notation
where $A_y^p(pd)$ is the proton analyzing power in $pd$ elastic scattering
and $A_y^p(pn)$ is the same in $pn$ elastic scattering. The deuteron vector
analyzing power in $dp$ elastic scattering is denoted by $A_y^d(dp)$.} is of
the form of $q$ times a function of $q^2$ and the ANKE 796~MeV results of
Fig.~\ref{pdAy_STT} are well described by
\begin{equation}
A_y^{p}(pd) = 0.4714q-0.0987q^3+0.0077q^5.
\label{fit}
\end{equation}
This form also reproduces very well the LAMPF data~\cite{IRO1983} provided
that it is multiplied by a factor of $1.021$. This 2\% difference is to be
compared with the $3\%$ precision in the beam polarization measurements with
EDDA~\cite{WEI2000} and the 2\% systematic uncertainty in the LAMPF beam
polarization~\cite{MCN1980,MCN1981}.

By using only the information provided by the STT, there was a symmetric
setup that is certainly preferable for the measurements of an analyzing
power. However, the left-right symmetry is broken when information from the
forward detector is required, as it is in the measurement of quasi-elastic
scattering, to which we now turn.

\section{Analysing power in quasi-elastic proton-neutron scattering}
\label{NN}

\begin{figure}[h]
\includegraphics[width=1.0\linewidth]{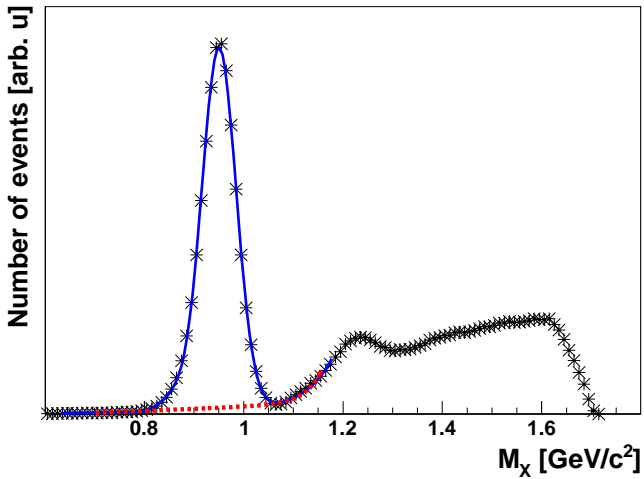}
\caption{Missing-mass $M_X$ spectra obtained for the $pd\to ppX$ reaction at
a beam energy of 1800~MeV when detecting one proton in the right side STT
and the other one in the FD. This distribution shows a clear neutron peak with
little background, estimated by the (red) dashed line of linear plus
exponential terms fitted using data from
outside the peak region. It is possible that the start of the continuum reveals
evidence for $\Delta(1232)$ excitation. The solid (blue) curve represents the
Gaussian + background fit to data in the neutron peak region.}
\label{missing_mass}
\end{figure}

Events corresponding to the breakup reaction $pd\to ppn$ can be identified by
measuring a fast proton in the forward detector and a slow one in one of the
STT. These then provide a missing-mass distribution for the $pp\to ppX$
reaction and this is illustrated in Fig.~\ref{missing_mass} for a beam energy
of 1800~MeV. The neutron peak is well separated from the inelastic continuum
and the background under the peak is only a few percent. Apart from the
ambiguities of the background, the $pd\to ppn$ events are fully reconstructed
so that it is possible to study regions of quasi-elastic $pn$ scattering.

Having identified the $pd\to ppn$ events, the next task is to isolate
quasi-free elastic $pn \to pn$ and, in particular, to remove contamination
from quasi-elastic scattering on the proton in the deuteron. This was first
studied in simulations of the $pd \rightarrow ppn_{\rm spec}$ and the $pd
\rightarrow pnp_{\rm spec}$ reactions within the framework of a simple
incoherent ``spectator'' model, which has been used successfully in the
measurement of spin observables with a polarized deuteron
beam~\cite{BYS1985}. The Fermi motion of nucleons in the deuteron was taken
into account using the Paris model~\cite{LAC1981} but, in the absence of
information at the higher energies, the differential cross sections for free
$pn$ elastic scattering was assumed to be equal to that of $pp$ except in the
Coulomb interaction region. The events generated were convoluted with the
ANKE acceptance using the GEANT program package~\cite{AGO2003}.

\begin{figure}[h!]
\includegraphics*[width=0.9\linewidth]{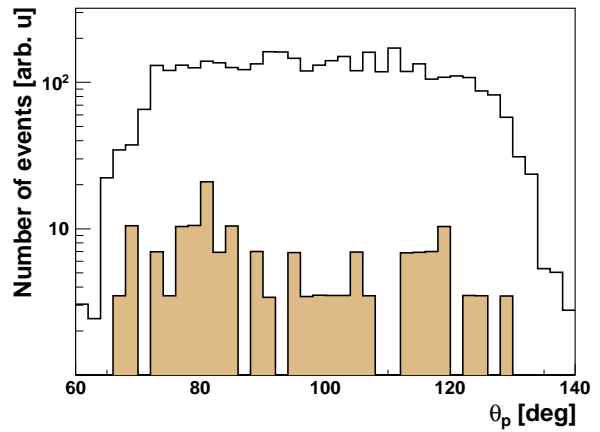}
\caption{Count rates for the $pd \rightarrow ppn_{\rm spec}$ (brown shading)
and $pd \rightarrow pnp_{\rm spec}$ (black lines) reactions simulated within
the framework of the incoherent ``spectator'' model at $T_p=800$~MeV. A fast
proton is detected in the FD and a slow one in the
right-side STT at a laboratory angle $\theta_p$.}
\label{simulation}
\end{figure}

As illustrated in Fig.~\ref{simulation}, even at the lowest beam energy, the
count rate from the $pd \rightarrow ppn_{\rm spec}$ reaction was found in the
simulation to be strongly suppressed kinematically compared with $pd
\rightarrow pnp_{\rm spec}$ when the slow proton was detected in the
right-side STT~\cite{BAR2013}. This is due to the asymmetric acceptance of
the FD and, for this reason, only data from the right-side STT were analyzed
in terms of quasi-elastic scattering on the neutron. This configuration also
reduces the contribution from the FSI between the ``spectator'' proton and
the recoil neutron. Despite the very simplified model used in the
simulations, the momentum and angular distributions for low energy protons in
events that formed the peak in Fig.~\ref{missing_mass} were found to be very
similar to the simulated distributions for ``spectator'' protons emitted from
the $pd \rightarrow pnp_{\rm spec}$ reaction. The count rate from quasi-free
$pp$ is expected in this model to be less than 5\% of that from quasi-free
$pn$ even at the 796~MeV. At higher beam energies the limit reduces to below
3\%.

Under the experimental conditions described above, the ANKE system operated
as a single-arm polarimeter, which means that the analyzing power had to be
deduced from the asymmetry of counts corresponding to different orientations
of the beam polarization. Such an asymmetry is very sensitive to the relative
luminosities of the beams with spin up and down. The ratio of luminosities
for $P^{\uparrow}$ and  $P^{\downarrow}$, integrated over a certain period of
data taking, was determined from the numbers of deuterons detected in both
STT during the same period, as described for elastic $\pol{p}d$ scattering in
sect.~\ref{pd}. If $|P^{\uparrow}| = |P^{\downarrow}|$ and the STT
acceptances were stable, the combination $(L_d^{\uparrow} \cdot
R_d^{\uparrow})/(L_d^{\downarrow} \cdot R_d^{\downarrow})$ would be equal to
the ratio of the squares of the luminosities, convoluted with the
``dead-time'' of readout system~\cite{OHL1973}. However, these conditions
should not be significant in our case. For example, because the $\pol{p} d$
asymmetry is less than 0.2 in our experiment, a 20\% difference between
$|P^{\uparrow}|$ and $|P^{\downarrow}|$ would induce a systematic effect in
$A_y^p(pn)$ that is below 1\%. A large difference in the STT acceptances for
different spin modes would also manifest itself in measurements of
$A_y^{p}(pd)$ presented in the previous section. Any significant effect can
be excluded here by comparing the 796~MeV ANKE and LAMPF data~\cite{IRO1983}
shown in Fig.~\ref{pdAy_STT}.

The ratio of luminosities obtained in this way could be unambiguously applied
for the normalization of the quasi-elastic data if these had been obtained
using the \emph{STT-trigger}. However, most of the \emph{STT-trigger} rate
was produced by particles that were accompanied by protons that did not fall
within the FD acceptance. In the more selective \emph{FdSTT-trigger}, a
coincidence was also demanded between a \emph{STT-trigger} signal and a
signal in the forward detector. Furthermore, in order to increase the number
of events recorded with the \emph{FdSTT-trigger}, the \emph{STT-trigger} rate
was significantly pre-scaled. Despite the whole ANKE detection system being
read out for any trigger, the ``dead-time'' corrections for data sets taken
with different triggers might still differ, and this has to be taken into
account. Nevertheless, it was found in a special investigation that the ratio
of the average ``dead-time'' factors obtained from data with beam
polarization up and down were nearly equal for both the \emph{STT-trigger}
and the \emph{FdSTT-trigger} data. The maximum deviation between the two
results was about 1\% but, on average, it was closer to 0.5\%.

The use of the ratio of luminosities derived from the numbers of deuterons
detected in both STT was also verified through the analysis of
proton-deuteron elastic events selected from data measured with both
triggers. Such a comparison was feasible because the \emph{STT-trigger} rate
was significantly pre-scaled so that it contained only a few percent of
events recorded with the \emph{FdSTT-trigger}. As stressed in the previous
section, the values of $A_y^{p}(pd)$ obtained using the cross-ratio method
are insensitive to the integrated luminosities convoluted with the
corresponding ``dead-time'' factors. In the case of the \emph{FdSTT-trigger},
the $\pol{p} d$ elastic events were selected by requiring the coincidence of
a proton detected in the FD with a deuteron identified in the left side STT.
The momentum of the fast proton was reconstructed in the same way as for $pn$
quasi-elastic events.

The angular dependence of the $pd$ elastic asymmetry derived from data
measured at 796~MeV with the \emph{FdSTT-trigger} is perfectly consistent
with that obtained in Ref.~\cite{IRO1983} and shown in Fig.~\ref{pdAy_STT}.
Furthermore, the average beam polarization of $0.502\pm0.002$, determined by
scaling our measured asymmetries to their analyzing powers, differs from the
value obtained with the EDDA polarimeter by only 2\%. At higher beam
energies, where no other measurements of the analyzing power have been found,
the asymmetry obtained from the \emph{FdSTT-trigger} data was compared with
that deduced from the \emph{STT-trigger} data using the cross-ratio method.
The results were found to be in good agreement in angular regions where there
was an overlap. A systematic difference of about 4\% was observed at
2157~MeV, though differences below 2\% were found at all the other energies.
These differences can be taken as estimates of the overall systematic
uncertainties when determining asymmetries with a single-arm polarimeter. In
addition to possible changes in acceptance for different spin modes, there
are also systematic uncertainties arising from possible differences between
beam polarizations $|P^{\uparrow}|$ and $|P^{\downarrow}|$ after
acceleration.

As shown in Fig.~\ref{missing_mass}, the background under the neutron peak at
1800~MeV was only about 6\%, and this was similar at other beam energies. If
the background analyzing power is large, it could nevertheless affect the
results because the $\pol{p}n$ asymmetry is typically about 0.1 or even less.
The background contribution was therefore evaluated for each angular bin and
the asymmetry corrected. When applying this correction, it was important to
ensure that the background was independent of beam polarization. For this
purpose, missing-mass distributions measured with $P^{\uparrow}$ and
$P^{\downarrow}$ for each energy were normalized to have equal luminosity and
then subtracted. For all the energies above 796~MeV the resulting
distributions contained only the neutron peak, which was very well fit by a
Gaussian distribution with no background. However, due to a small number of
deuterons originating from the $\pol{p}d \rightarrow p_{\rm spec} d \pi^0$
reaction, the background in the vicinity of the peak at the lowest beam
energy was found to depend on the polarization. After eliminating these
events by using the energy-loss information from the FD scintillation
hodoscope, the residual background was also shown to be polarization
independent. The systematic uncertainty arising from the description of the
background under the peak was estimated to be about $1.5\%$.

Taking into account the $3\%$ systematic error in the measurements of the
beam polarization with EDDA, we estimate that the overall systematic
uncertainty in the measurement of the analyzing power in $\pol{p}n$ elastic
scattering is about 5.5\% at 2157~MeV but below 4\% at the other beam
energies. These systematic effects are smaller than the typical statistical
errors of about 10\%.

\begin{figure}[htb]
\begin{center}
\includegraphics[scale=0.45]{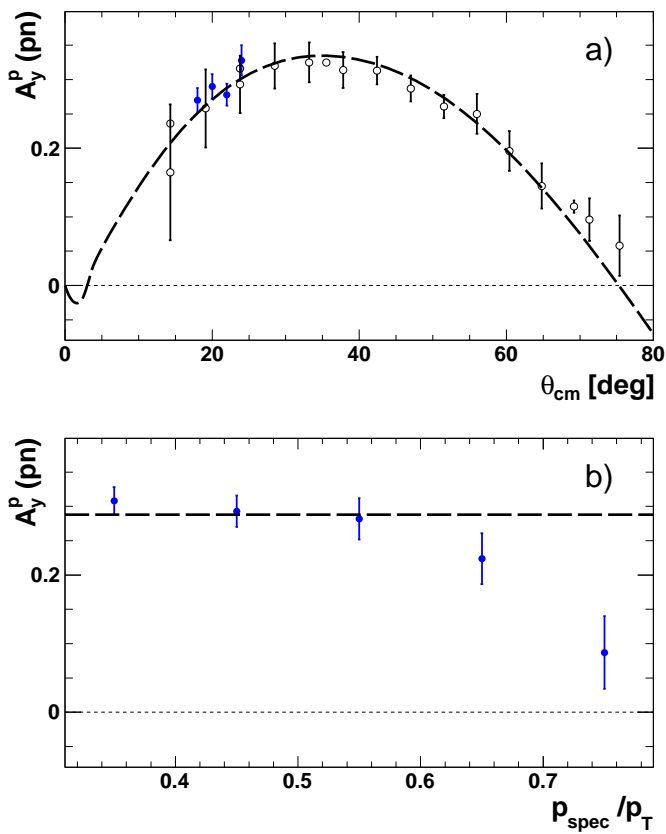}
\end{center}
\caption{Analyzing power $A_y^{p}(pn)$ in quasi-free $\pol{p} n$ elastic
scattering at $T_p \approx 796$~MeV. Panel~(a): The open points are results
from Ref.~\cite{BAR1983} as a function of the centre-of-mass scattering angle
$\Theta_{cm}$. The blue solid points show results from the current experiment
extracted, as discussed in the text, under conditions where $p_{\rm spec}/p_{T}
< 0.5$ and $p_{T}> 200$~MeV/$c$, where $p_T$ is the laboratory momentum
transfer. The predictions of the SAID SP07 partial wave
solution~\cite{ARN2007} are shown by the dashed curve. Panel~(b): The values of
$A_y^{p}(pn)$ measured at ANKE for $\Theta_{cm} = 22^{\circ}\pm2^{\circ}$ as a
function of the $p_{\rm spec}/p_{T}$ ratio. The dashed line indicates the SAID
SP07 solution for $A_y^{p}(pn)(\Theta_{cm}=22^{\circ})$~\cite{ARN2007}.
\label{pn0SAID}}
\end{figure}

In earlier experiments~\cite{BAR1983,DIE1975,MAK1980}, quasi-free
$A_y^{p}(pn)$ was studied by measuring both scattered particles in conditions
close to free kinematics and then reconstructing the momentum of the
unobserved ``spectator'' proton. In contrast, at ANKE the fast scattered
proton was detected in coincidence with the directly measured ``spectator''
proton. The quasi-free scenario is generally assumed to be realized when the
momentum transfer from a beam particle to a scattered one ($p_{T}$) is large
compared with the ``spectator'' particle momentum ($p_{\rm spec}$), which
should correspond to the Fermi momentum in the deuteron. It is clearly
desirable to determine experimentally the values of $p_{\rm spec}/p_{T}$ for
which the the ``spectator'' model is valid. This will be influenced by the
design of the STT, which requires a proton to have a momentum above
70~MeV/$c$ in order to be reconstructed.

The applicability of the ``spectator'' model was tested in the 796~MeV data.
Although only the laboratory momentum transfer range $100 < p_T <
260$~MeV/$c$ was here accessed, this was the only energy where several
experiments on quasi-free $\pol{p}p$ and $\pol{p}n$ elastic scattering were
performed~\cite{BAR1983,BAL1993,GLA1990,GLA1993} and which were used in the
derivation of the stable solution (SP07) of the SAID phase-shift
analysis~\cite{ARN2000,ARN2007}.

It is interesting to note that the $\pol{p}n$ analyzing power obtained
without any restriction on the $p_{\rm spec}/p_{T}$ ratio was found to be in
good agreement with the SP07 prediction over the whole of the ANKE angular
acceptance which, at this beam energy, is $10^{\circ} <\Theta_{cm}<
25^{\circ}$. However, it is difficult to believe that the ``spectator'' model
could be still valid when $p_{\rm spec}/p_{T} > 0.5$ as this corresponds to
$\Theta_{cm} < 15^{\circ}$, i.e., a region where the $pn$ final state
interaction is very strong. The dependence of the analyzing power on the
$p_{\rm spec}/p_{T}$ ratio was therefore investigated separately in different
$\Theta_{cm}$ ranges.

The results for $20^{\circ} < \Theta_{cm} <  24^{\circ}$, which correspond to
momentum transfers $200 < p_T < 260$~MeV/$c$, are presented in the lower
panel of Fig.~\ref{pn0SAID}. The minimum value of $p_{\rm spec}/p_{T}$
allowed by the ANKE setup at this energy is $0.3$ but the values obtained for
$A_y^{p}(pn)$ remain close to the SP07 prediction up to $p_{\rm spec}/p_{T}
\simeq 0.6$. Using the conservative upper limit of $p_{\rm spec}/p_{T} <
0.5$, values of the analyzing power were obtained that were in good agreement
with the SP07 solution as well as with the data measured in
Ref.~\cite{BAR1983} down to $\Theta_{cm}=17^{\circ}$, as shown in the upper
panel of Fig.~\ref{pn0SAID}\footnote{In order to improve the clarity of the
figure, data from other experiments are not presented here.}. However, for
angles smaller than $17^{\circ}$ the dependence of $A_y^{p}(pn)$ on $p_{\rm
spec}/p_{T}$ was less credible. The analyzing power near the lower limit of
$p_{\rm spec}/p_{T}$ allowed by the FD acceptance was found to be
unexpectedly larger than that predicted by the SP07 solution and it decreased
monotonically with increasing $p_{\rm spec}/p_{T}$. This means that the
analyzing power measured for $\Theta_{cm} < 17^{\circ}$ with the $p_{\rm
spec}/p_{T} < 0.5$ cut deviates significantly from the expected angular
dependence. This deviation can be ascribed to the final state interaction
between the recoiling neutron and proton, which increases in importance as
$p_T$ is reduced.

\begin{table*}[ht]
\centering
\begin{tabular}{|c|c|c||c|c|c||c|c|c|}
\hline
$T_p$ & $\Theta_{cm}$ & $A_y^{p}(pn)$&$T_p$ & $\Theta_{cm}$ & $A_y^{p}(pn)$&$T_p$ & $\Theta_{cm}$ & $A_y^{p}(pn)$ \\
MeV   & degrees  &   &MeV   & degrees  &   &MeV   & degrees  &   \\
\hline
           &  18.0    &     $0.270\pm0.018$ &         &  13.5    &  $0.130\pm0.023$&         &  13.5   &     $0.125\pm0.011$    \\
   796     &  20.0    &     $0.290\pm0.018$ & 1600    &  16.5    &  $0.151\pm0.016$& 1800    &  16.5   &     $0.147\pm0.009$    \\
           &  22.0    &     $0.278\pm0.017$ &         &  19.5    &  $0.153\pm0.015$&         &  19.5   &     $0.156\pm0.009$    \\
           &  24.0    &     $0.328\pm0.022$ &         &  22.5    &  $0.162\pm0.017$&         &  22.5   &     $0.149\pm0.009$    \\
           &          &                     &         &  25.5    &  $0.175\pm0.016$&         &  25.5   &     $0.163\pm0.010$    \\
\hline
\hline
           &  14.0    &     $0.115\pm0.022$ &         &  13.5    &  $0.081\pm0.020$&         &  17.5   &     $0.088\pm0.012$    \\
   1965    &  16.5    &     $0.125\pm0.013$ & 2157    &  16.5    &  $0.104\pm0.014$& 2368    &  20.5   &     $0.112\pm0.013$    \\
           &  19.5    &     $0.127\pm0.014$ &         &  19.5    &  $0.110\pm0.014$&         &  23.5   &     $0.107\pm0.015$    \\
           &  22.5    &     $0.130\pm0.015$ &         &  25.5    &  $0.120\pm0.018$&         &  27.0   &     $0.120\pm0.016$    \\
           &  25.5    &     $0.146\pm0.018$ &         &  28.0    &  $0.140\pm0.025$&         &         &                        \\
           &          &                     &         &  28.0    &  $0.140\pm0.025$&         &         &                        \\
\hline
\end{tabular}
\caption{Analysing power $A_y^{p}(pn)$ in $\pol{p}n$ quasi-elastic scattering
measured at six proton kinetic energies $T_{p}$.} \label{tab:Aynp}
\end{table*}

The value of $\Theta_{cm} = 17^{\circ}$ at 796~MeV corresponds to a momentum
transfer of 180~MeV/$c$ and the data at the various energies reported in
Table~\ref{tab:Aynp} were all obtained with the restriction $p_T >
190$~MeV/$c$ as well as $p_{\rm spec}/p_{T}< 0.5$. This value of $p_T$ is at
the lower edge of the momentum transfer range covered by the FD detector at
1600~MeV and at higher energies it is well outside the range and therefore
does not introduce extra cuts.

\begin{figure}[htb]
\begin{center}
\includegraphics[scale=0.4]{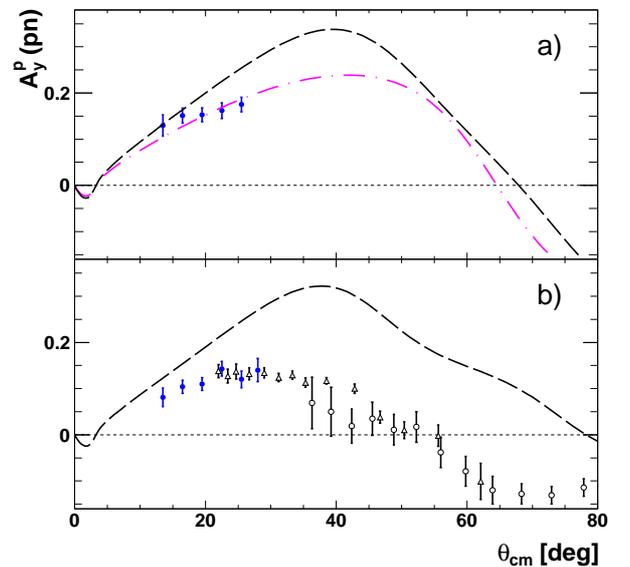}
\end{center}
\caption{$A_y^{p}(pn)$ of quasi-free $\pol{p} n$ elastic scattering at (a)
1600~MeV and (b) 2200~MeV as function of the centre-of-mass scattering
angle $\Theta_{cm}$. Solid (blue) points show the values obtained in the ANKE
experiment whereas open (black) points are results taken from
Ref.~\cite{DIE1975} (triangles) and Ref.~\cite{MAK1980} (circles). The magenta
dot-dashed curve represents the new
AD14 SAID solution at 1600~MeV~\cite{WOR2016}, though it should be noted that
this energy is outside the stated range of validity of this solution. The
dashed (black) curves in both panels illustrate the previous SAID
solution~\cite{ARN2007}, though it must be stressed that this also has limited
validity above 1300~MeV.\label{pnAsym}}
\end{figure}

The values of the analyzing power shown in Table~\ref{tab:Aynp} generally
decrease with increasing beam energy and the results presented in
Fig.~\ref{pnAsym} illustrate the scale of the dependence. Despite the
different experimental approach, the ANKE results at 2200~MeV are fully
consistent with data from Refs.~\cite{DIE1975,MAK1980}. As was stressed
already, the data base on $p n$ elastic scattering observables above 1500~MeV
is insufficient to yield reliable partial wave solutions. It is therefore not
surprising that the SAID SP07 solution~\cite{ARN2007} does not predict well
our new experimental data shown in Fig.~\ref{pnAsym}. However, the SAID
solution was recently updated to take into account the experimental data
measured at COSY-WASA~\cite{pnWASA}. Although it was asserted that the new
AD14 solution~\cite{WOR2016} is still valid only up to 1300~MeV, it,
nevertheless, gives predictions that are much closer to our 1600~MeV data
shown in Fig.~\ref{pnAsym} than those of SP07~\cite{ARN2007}.

\section{Conclusions}
\label{Conclusions}

We have measured the analyzing power in $\pol{p} d$ elastic and $\pol{p}n$
quasi-elastic scattering at 796~MeV and at five energies from 1600~MeV to
2400~MeV at the COSY-ANKE facility. The results at 796~MeV are consistent
with published data to within the quoted uncertainties. The $\pol{p}d$
elastic measurements at 1600~MeV and above were carried out for the first
time at small angles and there is little $\pol{p}n$ elastic information at
these higher energies.

The results on $\pol{p} d$ elastic scattering were obtained using two silicon
tracking telescopes as a two-arm polarimeter. In this case the systematic
uncertainty was mainly associated with the calibration of the EDDA beam
polarimeter, which is known with an accuracy of about $3\%$. Our results at
796~MeV lie about 2\% lower than the previous measurements~\cite{IRO1983} but
are easily consistent within the systematic uncertainties of both
experiments.

The analysing power in $\pol{p} d$ elastic scattering at higher energies was
found to be about a factor of two smaller than at the 796~MeV and generally
decreasing with beam energy. The decrease of analyzing power with energy is
similar to that noted for the deuteron analyzing power in $\pol{d}p$ elastic
scattering~\cite{HAJ1987,GHA1991,ARV1988,MCH2018}. This similarity is not
surprising because it has been argued in connection with the 796~MeV
data~\cite{IRO1983} that the proton analysing power at small angles is
determined mainly by the interference of the charge-average central $NN$
amplitude with the spin-orbit term. This should also be true for the deuteron
analyzing power, though there are of course different modifications of the
polarizations due to the multiple scatterings.

\begin{figure}[htb]
\begin{center}
\includegraphics[scale=0.4]{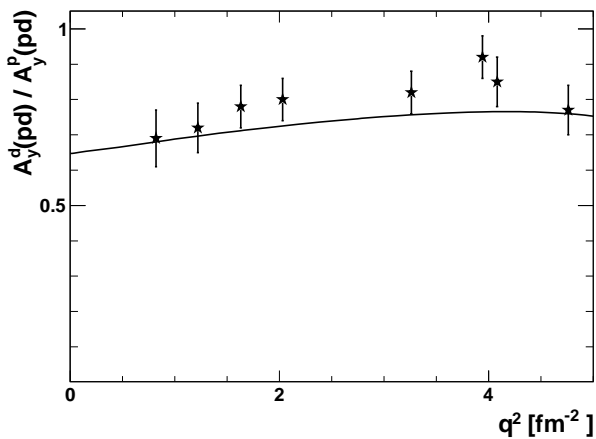}
\end{center}
\caption{The ratio between the vector analyzing power of the deuteron to that
of the proton in $pd$ elastic scattering at 796~MeV per nucleon. The values of
$A_y^{d}(dp)$ of Ref.~\cite{HAJ1987} have been read from a figure produced by
the same group~\cite{GHA1991}
whereas those of $A_y^{p}(pd)$ were taken from the fit of Eq.~(\ref{fit}) to the
current data. The curve represents the results of an extended multiple
scattering model using the formulae given in Ref.~\cite{PLA2010}. The results
are presented as a function of $q^2$ and the largest scattering angle shown
corresponds to $\Theta_{cm}\approx 30^{\circ}$.
\label{Ratio}}
\end{figure}

In the single scattering approximation the dominant $NN$ amplitudes, where
one neglects the spin-spin term, would suggest that the ratio $R =
A_y^{d}(dp)/A_y^{p}(pd)$ should be constant with a value of $2/3$.
Parameterizing all the $NN$ amplitudes using the SAID SP07 partial wave
solution~\cite{ARN2007} and including multiple scatterings in an extended
Glauber model~\cite{PLA2010} gives the curve shown in Fig.~\ref{Ratio}. It is
here compared to data extracted from Refs.~\cite{HAJ1987,GHA1991} combined
with the current results. Several systematic effects in the $NN$ input cancel
in the prediction of the analyzing power ratio. However, one is always left
with systematic uncertainties in the ratio arising from the measurements of
the deuteron (4\%) and proton (3\%) polarizations. Never\-theless, the
comparison shown in Fig.~\ref{Ratio} does suggest that the proton and
deuteron analyzing powers are strongly linked.

Since it is not possible to detect neutrons at ANKE, the analyzing power in
proton-neutron elastic scattering was studied in quasi-free conditions using
a deuterium target. This was accomplished by measuring the fast scattered
proton in the forward detector in coincidence with the low energy
``spectator'' proton from the $\pol{p}d \rightarrow pnp_{\rm spec}$ reaction
being measured in one of the silicon tracking telescopes. This scheme relies
completely on the simple ``spectator'' model. The validity of the empirical
``spectator'' approach with our kinematic cuts was tested by comparing our
result at 796~MeV with data from other
experiments~\cite{BAR1983,BAL1993,GLA1990,GLA1993} as well as with the SP07
SAID partial wave solution~\cite{ARN2000,ARN2007}. It seems from this that
the ``spectator'' model can be used if the $p_{\rm spec}/p_{T}$ ratio is
restricted to be below 0.5 and $p_T>190$~MeV/$c$. These criteria were then
applied in the analysis of our higher energy data. Good agreement was found
between our data at 2157~MeV and the results from other
experiments~\cite{DIE1975,MAK1980}, despite the different experimental
approaches. Systematic uncertainties of our results were estimated to be
about 5.5\% at this energy and about 4\% at others.

Just as for proton-deuteron elastic scattering, the analyzing power in
quasi-elastic $\pol{p}n$ scattering at higher energies is almost a factor of
two smaller than at 796~MeV. There is also a similar general decrease with
increasing beam energy. However, the analyzing power at high energy is
significantly less than that found in $\pol{p}p$ elastic
scattering~\cite{BAG2014}.

We are grateful to other members of the ANKE Collaboration for their help
with this experiment and to the COSY crew for providing such good working
conditions. Useful discussions took place with J.~Haidenbauer regarding the
extended Glauber calculations. This material is based upon work supported by
the Forsch\-ungszentrum J\"ulich (COSY-FEE) and the Shota Rustaveli National
Science Foundation Grant 09-1024-4-200.


\end{document}